%Tex file called: colpin2.tex
%Contents: Preprint, January 14,  1998.
%version: paper.
%
\def\gsim{\mathrel{\scriptstyle{\buildrel > \over \sim}}}
\def\lsim{\mathrel{\scriptstyle{\buildrel < \over \sim}}}
%\magnification=1095
\magnification 1200
\baselineskip=17pt
%\font\twelverm=cmr10 scaled 1200

%\vskip 32pt

\centerline{\bf CRITICAL BEHAVIOR OF LAYERED SUPERCONDUCTING}
\bigskip
\centerline{\bf FILMS IN PARALLEL MAGNETIC FIELD}
\vskip 50pt
\centerline{J. P. Rodriguez}
\medskip
\centerline{{\it Instituto de Ciencia de Materiales,
Consejo Superior de Investigaciones Cientificas,}}
\centerline{{\it Universidad Autonoma de Madrid,
Cantoblanco, 28049 Madrid, Spain}{\footnote*{Present address.}} {\it and}}
\centerline{\it Dept. of Physics and Astronomy,
California State University,
Los Angeles, CA 90032, USA.}
\vskip 30pt
\centerline  {\bf  Abstract}
\vskip 8pt\noindent
The equilibrium magnetization for layered superconducting films
that experience a nonzero component, $H_{\parallel}$, 
of magnetic field applied parallel to the layers
is computed at temperatures and at perpendicular field components
in the vicinity of the decoupling transition. 
A fermion analogy is exploited for this purpose,
whereby it is found that the parallel magnetization
shows an anomalous $H_{\parallel}^{-1}$ tail at high
fields due to entropic fluctuations of the
(parallel) lattice of  Josephson vortices.  A collective pinning effect 
is also identified for
$c$-axis transport limited by a single planar defect 
oriented parallel to both $c$ and to the applied magnetic
field.
%to this lattice.
% of Josephson vortices.
%$H_{\parallel}$.
\bigskip
\noindent
PACS Indices: 74.20.De, 74.20.Mn, 74.60.Ge, 74.80.Dm
\vfill\eject
 
\centerline{\bf I. Introduction}
\bigskip
	
It is well known that the electronic conduction in the normal state
of high-temperature superconductors (HTSC) is confined primarily
to the copper-oxygen planes common to these materials.$^1$
This suggests the Lawrence-Doniach (LD) model of a stack of
Josephson-coupled superconducting layers below $T_c$.$^{2,3}$
The following question can then be posed:$^4$
Is the superconducting phase strictly three dimensional (3D),
or does  two-dimensional (2D) superconductivity appear
somewhere in the phase diagram?  The experimental answer
is complex.  Conductivity measurements give evidence
for a 2D superconducting transition,$^5$  whereas thermodynamic measurements
are more consistent with a 3D transition.$^6$
The  resolution to this  paradox  is
that, although the thermodynamic transition into the superconducting
phase is in fact 3D,$^{7-10}$  quasi-2D behavior arises
for physics on length scales small in comparison to the Josephson
penetration length,$^{11}$ $\lambda_J$.
Indeed, the $c$-axis Josephson  plasma
resonance seen  
%at frequencies of order $\omega_{pl}\sim c/\lambda_J$ 
in HTSC for magnetic
fields oriented  perpendicular to the layers that exceed  the 2D-3D
cross-over scale,$^{12}$ $B_*^{\perp}\sim\Phi_0/\lambda_J^2$,
is an experimental example  of such 
behavior,$^{13}$   in which case Josephson coupling persists
in between layers while $c$-axis vortex lines have degenerated into
decoupled pancake vortices.$^3$

In this paper, we shall obtain  the critical properties
theoretically expected for thin films of
extreme type-II layered superconductors in magnetic field
aligned parallel to the layers.
The quasi-2D regime cited above is presumed throughout. 
This guarantees that the thermodynamics ``factorizes''
into that associated with (perpendicular)
pancake vortices and to that associated with
(parallel) Josephson vortices.$^{9, 14}$
In addition, we shall assume that the thickness of the
film (including leads) is less than the London
penetration length to insure the absence of magnetic screening across
the layers.
The theoretical analysis exploits a new fermion
analogy for the LD model.$^{15}$  We first compute the
parallel equilibrium magnetization.  It is found to exhibit
a series of renormalized melting transitions connected with the
parallel vortex lattice,$^{16,17}$
as well as an anomalous $H_{\parallel}^{-1}$ tail at high parallel
fields,$^{15}$ $H_{\parallel}$ (see Figs. 1 and 2).
We then compute $c$-axis transport limited by the presence
of a single planar defect or  pair of edges oriented parallel to both
the $c$-axis and  to the applied magnetic field.
A collective pinning regime is identified
in  the high-field limit (see Fig. 1) for planar pins   
with an effective thickness $\xi_p > a_0$, where $a_0$ denotes
the lattice constant of the parallel vortex lattice.
In particular, the
current-voltage ($I$-$V$)
characteristic is found to be {\it algebraic} in this case,
and to  be inversely related to that obtained
for a Luttinger liquid in the presence of a single backscattering
impurity.$^{18}$

\bigskip
\bigskip
\centerline{\bf II. Parallel Equilibrium Magnetization}
\bigskip

The problem to be solved then is the {\it parallel}
thermodynamics of a finite number, $N$ of Josephson coupled
planes in the quasi-2D regime$^{11-14}$
$H_{\perp} >  \Phi_0/\lambda_J^2$ and/or
$T > T_{cr}$, where the 2D-3D cross-over temperature $T_{cr}$ 
marks the point at which the intra-plane phase coherence
length, $\xi$, matches the Josephson penetration length, $\lambda_J$.
It is implicit then that  long-range in-plane phase
coherence is lost at a temperature $T_c$ below $T_*$,
at which point
the planes entirely decouple;$^{7-9}$
i.e., $\xi(T) = \infty$ for $T < T_c$, while
$\lambda_J(T) = \infty$ for $T > T_*$.
Since both $\xi$ and $\lambda_J$ are finite in the interval between
$T_c$ and $T_*$, the relationship $\xi(T_{cr})\sim\lambda_J(T_{cr})$
then implies the inequalities
$T_c < T_{cr} < T_*$.$^{11}$
We shall now introduce the corresponding LD free-energy
functional, which reads
$$\eqalignno{
E_{\rm LD} =
J_{\parallel}\int d^2 r \Biggl[
\sum_{l = 1}^{N}
{1\over 2}(\vec\nabla\theta_l)^2
-\Lambda_0^{-2}
\sum_{l = 1}^{N-1}{\rm cos}(\theta_{l+1}
-\theta_l -A_z)\Biggr]\cr
&&(1)\cr}$$
in the absence of magnetic screening.  Here, 
$\theta_l(\vec r)$ denotes the phase of the superconducting
order parameter in layer $l$, where $\vec r = (x, y)$
is the planar coordinate.  The parallel magnetic
induction $B_{\parallel} = (\Phi_0/2\pi d) b_{\parallel}$
found between layers $l$ and $l+1$ and aligned along
the $y$ axis is related to the vector potential
above by $A_z = - b_{\parallel} x$.  Here $d$ represents
the spacing between layers.  Last, $J_{\parallel}$ is a measure
of the local in-plane phase rigidity, while $\Lambda_0$
sets the bare scale for the Josephson penetration  length.
The author has recently obtained as analogy
between the  above LD model and
coupled chains of spinless fermions at zero temperature,
where each chain corresponds to a layer.$^{15}$
Specifically, the Hamiltonian for the fermion
model   is divided into two parts, 
$H = H_{\parallel} + H_{\perp}$, with
%$$\eqalignno{
%H_{\parallel} =  \sum_{l = 1}^N\int dx\Biggl\{ 
%v_F \Bigl[\Psi_L^{\dag}(x,l) & i\partial_x \Psi_L(x,l) 
%- \Psi_R^{\dag}(x,l) i\partial_x \Psi_R(x,l)\Bigr]\cr
%%-\mu_l\Bigl[\Psi_L^{\dag}(x,l) \Psi_L(x,l) 
%%+ \Psi_R^{\dag}(x,l) \Psi_R(x,l)\Bigr]\Biggr\}\cr
%& + U_{\parallel}\Psi_L^{\dag}(x,l)\Psi_R^{\dag}(x,l)
%\Psi_L(x,l)\Psi_R(x,l)\Biggr\} & (2a)\cr}$$
$$
H_{\parallel} =  \sum_{l = 1}^N\int dx\Biggl[ 
v_F \Bigl(\Psi_L^{\dag}  i\partial_x \Psi_L
- \Psi_R^{\dag} i\partial_x \Psi_R\Bigr)
 + U_{\parallel}\Psi_L^{\dag}\Psi_R^{\dag}
\Psi_L\Psi_R\Biggr]\eqno (2a)$$
and
$$\eqalignno{
H_{\perp} = & U_{\perp} \sum_{l=1}^{N-1}\int dx
\Bigl[\Psi_L^{\dag}(x,l)\Psi_R^{\dag}(x,l+1)\Psi_L(x,l+1)\Psi_R(x,l)
+ {\rm H.c.}\Bigr], & (2b)\cr}$$
and with field operators 
$\Psi_R(x,l)$ and $\Psi_L(x,l)$ for right ($R$) and left
($L$) moving fermions.
The coordinate along the Josephson vortices, $y$,
is related to the imaginary time
variable $\tau$ of the fermion analogy by
$y = v_F^{\prime}\tau$. 
Here, the Fermi velocity $v_F^{\prime} = v_F\, {\rm sech} \, 2\phi$
is renormalized by the intra-chain
interaction $U_{\parallel}$,$^{19}$
with  ${\rm tanh}\, 2 \phi = U_{\parallel}/2\pi v_F$. 
% sets the phase $\phi$. 
Also, $U_{\perp} > 0$ is a repulsive backscattering 
interaction energy$^{19}$
in between chains.
The Gibbs free-energy of the LD model (1) 
with respect to the normal state is then found to be
related to the ground-state energy $E_F$ of the fermion analogy by 
$$(G_s - G_n)/k_B T = (L_y/v_F^{\prime}) [E_F(U_{\perp}) - E_F(0)].
\eqno (3)$$
The identifications
$$\eqalignno{
b_{\parallel} = & 2\pi(N_{l+1} - N_{l})/L_x, & (4)\cr
T = &  e^{2\phi} T_{*0}, & (5)\cr
\Lambda_0^{-2} = & \alpha^{-2} (2|U_{\perp}|/\pi v_F^{\prime})
(T/T_{*0}),
& (6)\cr}$$
complete the equivalence between the models, where $N_l/L_x$
gives the fermion density in the  $l^{\rm th}$ chain, 
%$T_{dp} = 2\pi J_{\parallel}$ is the depinning temperature (see below), 
$T_{*0} = 4\pi J_{\parallel}$ is the decoupling temperature that
marks the point at which interlayer phase coherence is lost,$^{7-9}$
and where $\omega_0 = v_F^{\prime}\alpha^{-1}$ is the ultraviolet
cutoff in energy.  
This analogy is a direct generalization of the well-known equivalence 
that exists between the sine-Gordon model and the massive 
Thirring/Luther-Emery model 
in $1+1$ dimensions
to a layered structure.$^{21}$
%The above fermion analogy 
%is known to be exactly soluble
It reduces to a free theory
in the double-layer case ($N=2$) along the Luther-Emery (LE) line
$T = 2\pi J_{\parallel}$.$^{19}$

The general case, however, can be treated in the
mean-field  approximation defined by the charge-density wave (CDW)
order parameter
$\chi_l(x) = \langle\Psi_R^{\dag}(x,l)\Psi_L(x,l)\rangle$, and the
associated gap equation$^{22}$
$$\Delta_l = U_{\parallel}\chi_l + U_{\perp}
(\chi_{l+1} +\chi_{l-1}).\eqno (7)$$
[The order parameters at the boundaries are set
to $\chi_0(x) = 0 = \chi_{N+1}(x)$.]
The mean-field Hamiltonian then has the form
$H_{\rm MF} =  \sum_{l=1}^N \int dx
\Psi_l^{\dag} (H_l - \mu_l) \Psi_l$,
where the  spinor field,
$\Psi_l (x) = (\Psi_L(x,l), \Psi_R(x,l))$, for each layer
is acted upon by the
one-body operator
$$H_l =   \sigma_3 v_F i\partial_x + \sigma_+ \Delta_l(x)
+ \sigma_- \Delta_l^*(x).\eqno (8)$$
Here we define  $\sigma_{\pm} = {1\over 2}(\sigma_1 \pm i\sigma_2)$,
where  
$\sigma_i$ represent the Pauli matrices.  Also,
$\mu_l$ denotes the chemical potential for the fermions in layer $l$.
At zero parallel magnetic induction, the mean-field ground
state has spin-density wave (SDW) type order,
$\chi_l (x) = (-1)^l \chi_0$, for $U_{\perp} > 0$.
In  the minimal case, $N=2$, a constant gap 
$\Delta_0  = \omega_0 / {\rm sinh}[2\pi v_F/(U_{\perp}- U_{\parallel})]$ 
opens at each Fermi surface
for interaction parameters
satisfying $U_{\perp} >  U_{\parallel}$.
The latter phase boundary agrees with renormalization
group results.$^{23}$
Also, by  Eqs. (3) and (4), the line-tension of a single Josephson vortex
is given in general by
$\varepsilon_{\parallel} = k_B T |\Delta_{\sigma}|/v_F^{\prime}$,
with a pseudo-spin gap equal to
$\Delta_{\sigma} = \Delta_0(U_{\perp}) - \Delta_0(0)$.
For weak coupling, $U_{\perp}\rightarrow 0$,
we then  have $\Delta_{\sigma} =
U_{\perp}{\partial\over{\partial U_{\perp}}}\Delta_0|_{U_{\perp}=0}$
at temperatures $T < T_{*0}$
(or $U_{\parallel} < 0$),
while $\Delta_{\sigma} = \Delta_0$ at  temperatures
$T > T_{*0}$ (or $U_{\parallel} > 0$).
The former yields  a pseudo-spin gap  of
$\Delta_{\sigma} \cong 0.9 (U_{\perp}/2\pi\alpha)$
along the LE line $T = 2\pi J_{\parallel}$
for the case $N=2$, 
which is comparable
to the exact value of
$\Delta_{\sigma} = U_{\perp}/2\pi\alpha$.$^{19}$ 
The success that this mean-field approximation has in the double-layer case
indicates that it is reliable for the case of large $N$, where
fluctuations are smaller.
A constant gap,
$\Delta_0  = \omega_0 / {\rm sinh}[2\pi v_F/ (2U_{\perp}- U_{\parallel})]$, 
opens at each Fermi surface in such case
%, $\mu_l = v_F k_{F,l}$, 
for interaction parameters
satisfying $U_{\perp} > {1\over 2} U_{\parallel}$.
The identification (6) then implies the leading
dependence $T_* = 4\pi J_{\parallel}[1+{1\over 2}(\alpha/\Lambda_0)^2]$
of the decoupling temperature with the bare
Josephson scale.
Also, the pseudo-spin gap now has value
$\Delta_{\sigma} \cong 1.9 (U_{\perp}/2\pi\alpha)$
along the LE line. %, $T = 2\pi J_{\parallel}$,
The parallel lower critical
field, $H_{c1}^{\parallel} = 4\pi\varepsilon_{\parallel}/\Phi_0$,
vanishes exponentially at the decoupling temperature $T_*$, however, 
since $\Delta_{\sigma}\propto\Delta_0(T)$.   This implies an inflection
point in its temperature profile below
$T_*$.$^{14,24}$

%\bigskip
%\centerline{\bf A. Low-Field Limit}
%\bigskip
      
The presence of a parallel magnetic field, however, will
generally induce  the phase 
$\theta_l^{\prime}(x)$ 
of the CDW
order parameter, 
$\chi_l = (-1)^l\chi_0 \,{\rm exp}(i\theta_l^{\prime})$,
to wind.  This can be seen explicitly from the Ginzburg-Landau
(GL) equations that describe the present mean-field theory
at ``temperatures'' $T^{\prime} = v_F^{\prime}/L_y$
near $T_c^{\prime}\sim \Delta_0$  [see Appendix A, Eq. (A10)].
After extending Gorkov's original derivation$^{25}$ of the GL
equations to  the case of an isolated pair
($l, l+1$) of consecutive layers, 
one recovers the original LD free energy functional (1), 
%in the vicinity of the LE line 
%$^{26}$
but with  the bare
scale $\Lambda_0$ replaced by the renormalized
Josephson penetration length 
$\lambda_J = v_F/|\Delta_{\sigma}|$.$^{14}$
This is a result of
entropic wandering of the
Josephson vortices in the parallel direction, which
is particularly relevant in the regime of low
parallel fields, $b_{\parallel} \ll \Lambda_0^{-1}$.
Note that the  later corresponds precisely  to
the long wave-length limit in which Gorkov's derivation
of the GL equations is valid.$^{25}$
Also, the CDW phase is related to the true phase in the LD
model (1) by the gauge transformation
$\theta_l =  \theta_l^{\prime} - 2 k_{F,l} x$, where
$k_{F,l}$ denotes the Fermi wave-vector of layer $l$.
The  formula 
$N_l = \pi^{-1} k_{F,l} L_x$ then indicates
that the number of fermions, $N_l$,  is equal to the winding number
of the CDW phase
in a given chain $l$.  This fact
coupled with the identification
(4)  also  demonstrates that the gauge-invariant phase difference
between these  layers is just
$\theta_{l+1} -\theta_l - A_z = \theta_{l+1}^{\prime}-\theta_l^{\prime}$.
A  GL theory analysis$^{16}$ of the {\it renormalized}
LD model$^{14}$ (1) then yields a sequence of first-order
commensuration transitions 
at parallel fields $H_{l_1}^{\parallel}$ between
a  parallel vortex lattice with flux penetration every 
$l_1$ layers to one with flux penetration every $l_1+1$
layers as the field is lowered.  Fig. 1  displays
the predicted phase diagram in the critical regime 
using the  estimates
$H_1^{\parallel} = {1\over 3} \Phi_0/\lambda_J d$
and $H_2^{\parallel} = {3\over 8} H_1^{\parallel}$
based on the continuum limit. 
Contrary to claims in the literature,$^8$ we therefore find
no evidence for a true layer decoupling transition as a
function of parallel field in the critical regime.$^{14,15}$

%\bigskip
%\centerline{\bf B. High-Field Limit}
%\bigskip

Consider now a parallel vortex lattice 
in the limit of high parallel field,
but where  flux penetrates only every $l_1 \geq 2$ layers. 
The LD model (1) indicates
that spatial variations of the superconducting phase are generally
absent in the former limit;$^{14,16}$ i.e., 
$\theta_l(x) = 0$ and $\theta_l^{\prime}(x) =  2 k_{F,l} x$.
If there exists {\it no} flux penetration in between layers
$l$ and $l\mp 1$, the gap function (7) then
has the form
$$\Delta_l = (-1)^{l\pm 1} U_{\perp} \chi_0 
e^{i(\theta_{l\pm 1}^{\prime} - \theta_{l}^{\prime})}
= (-1)^{l+1}\Delta_{\sigma} e^{\pm i b_{\parallel}x}\eqno (9)$$
along the special line $U_{\perp} = U_{\parallel}$
in parameter space, where $b_{\parallel}$ denotes the
flux in between layers $l$ and $l\pm 1$
and where $\Delta_{\sigma} = U_{\perp}\chi_0$.
Here, we have made a  gauge transformation in order
to set  
$\mu_l$ and $k_{F,l}$ to zero (see ref. 22).  The
mean-field Hamiltonian 
%$H_l$ [see Eq. (8)]  
(8) then has energy eigenvalues
$$\varepsilon_k^{\pm} = 	 v_F k_F
\pm [v_F^2(k - k_F)^2 +\Delta_{\sigma}^2]^{1/2},
\eqno (10)$$
with an effective Fermi wave number
$k_F = \pm b_{\parallel}/2$ and a pseudo-spin gap
$\Delta_{\sigma}$. 
%= \Delta_0$.
The equilibrium  magnetization 
$M_{\parallel} = -{\partial\over{\partial H_{\parallel}}}
[(G_s-G_n)/V]$
can then be computed using the equivalence (3).
After  following
steps similar to those taken in the double-layer
($l, l\pm 1$)  case along the
LE line,$^{15}$ one obtains the formula
$$-4\pi M_{\parallel}^{(l_1)} = {1\over 2} H_{c1}^{\parallel}
\Biggl\{\Biggl[1+
\Biggl({H_{\parallel}\over{l_1^{-1}B_*^{\parallel}}}\Biggr)^2\Biggr]^{1/2} -
{H_{\parallel}\over l_1^{-1} B_*^{\parallel}}\Biggr\}\eqno (11)$$
for the equilibrium parallel magnetization,
where $B_*^{\parallel} = \Phi_0/\pi\lambda_J d$ is the
parallel field cross-over scale, with Josephson penetration
length $\lambda_J = v_F/|\Delta_{\sigma}|$.  Given the phase
diagram (Fig. 1) arrived at by the previous low-field analysis,
this result implies a series of jumps
{\it down} in $M_{\parallel}$ 
at each melting field, $H_{l_1}^{\parallel}$,
as parallel field increases (see Fig. 2).  Notice
that the sign of the jump in this case is opposite to
that observed in HTSC for vortex lattice melting
as the perpendicular magnetic field 
is swept.$^{26}$
Last, (9) indicates that the gap function 
$\Delta_l(x)$ simply acquires a constant
phase factor upon a uniform translation
$x\rightarrow x + a_l$ of the coordinate in layer $l$.  
Hence, the parallel vortex lattice is
infinitely smectic in the high-field limit.  (This result is consistent
with the shear instability obtained in the standard GL analysis of
LD model (1).$^{3}$)
The parallel magnetization in the high-field limit ($l_1 = 1$) is then
given by that of an isolated double layer,$^{15}$
which coincides with (11) up to prefactor of $2^{1/2}$.
This means that  the parallel magnetization must show
an anomalous $H_{\parallel}^{-1}$ tail  at high fields.  

Finally, to demonstrate that the above formula (11) for
the parallel magnetization  in the high-field limit is
generic, we shall now analyze
the spectrum of the mean-field Hamiltonian (8) 
along another special line, $U_{\parallel} = 0$,
in which case 
degenerate perturbation theory in powers of $U_{\perp}$
can be employed.   Let us suppose again that 
flux penetrates only every $l_1 \geq 2$ layers,
and that there exists {\it no} flux penetration in between layers
$l$ and $l\mp 1$.
%$^{26}$
After performing the same gauge transformation as before to
set $k_{F,l}$ and $\mu_l$ to zero (see ref. 22), we
obtain a gap function (7) of the form
$\Delta_l = (-1)^{l+1}\Delta_{\sigma} (1 + e^{\pm i b_{\parallel}x})$,
where $b_{\parallel}$ denotes the
flux in between layers $l$ and $l\pm 1$, and where
$\Delta_{\sigma} = U_{\perp}\chi_0$.
Now in the absence of Josephson coupling, $U_{\perp} = 0$,
the mean-field Hamiltonian (8) has eigenstates
$\Psi_a = (0, e^{ikx})$ and $\Psi_b  = (e^{i(k - 2 k_F)x},0)$,
with the corresponding energy eigenvalues
$\varepsilon_a =  v_F k$ and  $\varepsilon_b = v_F (2k_F - k)$.
As before, we have $k_F = \pm b_{\parallel}/2$.
The application of degenerate perturbation theory in powers of $U_{\perp}$ 
with respect to such states at momenta $k\sim k_F$
then yields the well-known
formula
$$\varepsilon_k^{\pm} = {1\over 2} (\varepsilon_a + \varepsilon_b)
\pm\Bigl[{1\over 4}(\varepsilon_a - \varepsilon_b)^2
+ \Delta_{\sigma}^2\Bigr]^{1/2}$$
for the perturbed energy eigenvalues.
Substitution of the unperturbed energies above then yields the
previous result (10).
We thereby recover the formula (11) for the parallel magnetization
in the case that $l_1 \geq 2$.
Last, when the parallel flux penetrates all layers ($l_1 = 1$), the gap
function (7) has the form
$\Delta_l = (-1)^{l+1}\Delta_{\sigma}
(e^{\mp i b_{\parallel}x} + e^{\pm i b_{\parallel}x})$.
This means that we must take into account the previous  unperturbed states,
as {\it well} as their time-reversed counterparts
obtained after making the (global) replacement
$k_F\rightarrow - k_F$.  The end result is again the previous
formula (11) for the parallel magnetization in the case
$l_1 = 1$.

\bigskip
\bigskip
\centerline{\bf III. Parallel Collective Pinning}
\bigskip

We shall now compute $c$-axis transport limited by a single planar defect
or pair of edges in the high-field limit,
$H_{\parallel} > H_1^{\parallel}$, where parallel flux
penetrates in between every layer.  
The planar pin is assumed to be parallel to both the $c$-axis and
to the applied magnetic field.
Due to the extreme smecticity symptomatic of
this regime,   it is sufficient to study the double-layer
case.
The LD model (1) then reduces to a pinned sine-Gordon system$^{14}$
with free-energy functional
$$\eqalignno{
E_{\rm SG} = J_{\parallel}\int d^2 r 
\Bigl [{1\over 4}(\vec\nabla\theta_- - \hat x b_{\parallel})^2   
-\Lambda_0^{-2} {\rm cos}\, \theta_-\Bigr ] &
- \varepsilon_{p}^{\parallel} \int dy\,  {\rm cos}\, \theta_-|_{x_0}\cr
& + \varepsilon_{\rm L}^{\parallel} \int d^2 r\theta(x-x_0)
{\partial\theta_-\over{\partial x}}, & (12) \cr}$$
where $\theta_-$ is the gauge-invariant phase difference in between 
the consecutive layers.  Here,
$\varepsilon_{\rm L}^{\parallel} = c^{-1} (I/L_y) (\Phi_0/2\pi)$
is the line tension
due to the Lorentz force in
the absence of flux flow. 
The  Lorentz force
and the pinning force are in equilibrium in such case,
which implies
that  the $c$-axis current $I$ flows {\it only} at the
pin site $x_0$.$^{27}$
Also, since Josephson coupling is weaker
near the pin, we have that  $\varepsilon_{p}^{\parallel} < 0$.
Following Coleman$^{21}$ and LE,$^{19}$ this model is equivalent
to the massive Thirring model for 1D fermion fields, 
$\Psi = (\Psi_L, \Psi_R)$,
in the presence of a single  backscattering impurity.$^{18}$
Its Hamiltonian description is then
$$\eqalignno{
H_{\sigma} = 
\int dx \Psi^{\dag}
(\sigma_3 v_F^{\prime} i\partial_x 
+ \sigma_1 \Delta_{\sigma} + 2^{1/2} V_{\rm KF})
\Psi  &
+ 2 g_0 \int dx \Psi_L^{\dag} \Psi_R^{\dag}
\Psi_R \Psi_L \cr
& - \xi_p\Delta_{\sigma}
(\Psi_L^{\dag}\Psi_R + \Psi_R^{\dag}\Psi_L)|_{x_0}, & (13)\cr}$$
where $V_{\rm KF} (x) = 
2\pi v_F^{\prime} (\varepsilon_{\rm L}^{\parallel}/k_B T)
\theta (x - x_0)$ is the voltage drop equivalent 
to the
Lorentz force located at $x_0$,
while  $\xi_p\propto -\varepsilon_{p}^{\parallel}$
gives the effective thickness of the pin plane.  The interaction between
fermions is related to the depinning temperature (or LE line)
$T_{dp} = 2\pi J_{\parallel}$ by the relationship
$${T_{dp}\over T} = 1 + {g_0\over{\pi v_F^{\prime}}}.\eqno (14)$$ 
Notice then that the fermions
interact repulsively for $T < T_{dp}$, while
they interact attractively for $T > T_{dp}$.  Last, Eq. (4)
indicates that the total number of fermions is equal to the total
number of Josephson vortices lying in between the  consecutive
layers.  
 
Yet Eq. (14) indicates that  the Thirring model is nearly free at  temperatures
near $T_{dp}$,  with
a Fermi surface at $k_F = {1\over 2} b_{\parallel}$.
This suggests a canonical transformation of the form
$$\eqalignno{
c_k & =  x_k a_k + y_k b_k & (15)\cr
d_k & = -y_k a_k + x_k b_k & (16)\cr}$$
%After making a canonical transformation 
%a Luttinger liquid model
%with renormalized Fermi velocity and local backscattering term is
%obtained.
%$^{26}$  
to remove the (mass) gap term in (13) at momenta $k$ in the
vicinity of the Fermi surface.  Here,  the original field
operators for
right and left moving spinless fermions are represented as
$\Psi_R (x) = L_x^{-1/2}\sum_k e^{ikx} a_k$ and
$\Psi_L(x) = L_x^{-1/2}\sum_k e^{ikx} b_k$, respectively.
The former is achieved by the choice of coherence factors
$(x_k, y_k)  = (u_k, v_k)$  for $k>0$ and
$(x_k, y_k) =  (v_k, - u_k)$ for $k<0$, with
$$\eqalignno{
u_k & = 2^{-1/2} \Bigl(1 + {v_F^{\prime} k\over{E_k}}\Bigr)^{1/2}, & (17)\cr
v_k & = 2^{-1/2} \Bigl(1 - {v_F^{\prime} k\over{E_k}}\Bigr)^{1/2}, & (18)\cr}$$
along with energy eigenvalue
$$E_k = (v_F^{\prime 2} k^2 + \Delta_{\sigma}^2)^{1/2}.\eqno (19)$$
Notice then that the  energy eigenvalue for the new
right-moving state state $c_k^{\dag}|0\rangle$
is equal to $\varepsilon_k = \theta (k) E_k - \theta (-k) E_k$, while
it is equal to $-\varepsilon_k$ for the new
left-moving state $d_k^{\dag}|0\rangle$.
Finally, the field operators obtained after such a canonical
transformation are then
$\Psi_+ (x) = L_x^{-1/2}\sum_k e^{ikx} c_k$ and
$\Psi_-(x) = L_x^{-1/2}\sum_k e^{ikx} d_k$.
After making   the Luttinger liquid hypothesis, 
which assumes that only excitations near the Fermi surface are
relevant,
we obtain the  following effective {\it massless}
Thirring model Hamiltonian: 
$$\eqalignno{
H_{\sigma}^{\prime} =
\int dx \Psi^{\prime\dag}
(\sigma_3 v_F^{\prime\prime} i\partial_x
 + 2^{1/2} V_{\rm KF})
\Psi^{\prime}  &
+ 2 g_0^{\prime} \int dx \Psi_-^{\dag} \Psi_+^{\dag}
\Psi_+ \Psi_-\cr
& - \xi_p^{\prime}\Delta_{\sigma}
(\Psi_-^{\dag}\Psi_+ + 
\Psi_+^{\dag}\Psi_-)|_{x_0}, & (20)\cr}$$
where $v_F^{\prime\prime} = (v_F^{\prime}k_F/E_{k_F}) v_F^{\prime}$,
$g_0^{\prime} = (v_F^{\prime}k_F/E_{k_F})^2 g_0$
and $\xi_p^{\prime} = (v_F^{\prime}k_F/E_{k_F})\xi_p$ are the renormalized
Fermi velocity, interaction and pinning scale, respectively, and where
$\Psi^{\prime} = (\Psi_-, \Psi_+)$ is the
canonically transformed spinor field.
The above effective Luttinger model is valid in the limit of
({\it i}) a thick pinning plane,
$\xi_p > a_0$, with respect to the
separation in between Josephson  vortices, in the limit 
of ({\it ii}) high parallel
fields  $B_{\parallel} > \Phi_0/\lambda_J d$, and
at ({\it iii}) temperatures in the vicinity of the depinning
transition, $T\sim T_{dp}$ [see Appendix B, Eqs. (B1) and (B14)].
Notice that the sheer  existence of
the  gapless fermion analogy (20) indicates algebraic long-range order,
$\langle e^{i\theta_-(0)} e^{-i\theta_-(r)}\rangle\propto
r^{-2K_{\sigma}^{\prime}}$, for the vortex
lattice,$^{20}$ where 
$K_{\sigma}^{\prime} = [2(T/T_{dp}) - 1]^{1/2}$ at
high fields.
Finally, transforming {\it back} the massless Thirring model
(20)  to the bosonic description, 
one recovers the gaussian limit 
of the sine-Gordon model (12) in zero parallel
field, 
$$E_{\rm LL} = J_{\parallel}^{\prime}\int d^2 r
{1\over 4}(\vec\nabla\theta_-)^2
- \varepsilon_{p}^{\parallel} \int dy\,  {\rm cos}\, \theta_-|_{x_0}
 + {v_F^{\prime}\over{v_F^{\prime\prime}}} 
\varepsilon_{\rm L}^{\parallel} \int d^2 r\theta(x-x_0)
{\partial\theta_-\over{\partial x}},\eqno (21)$$
%[1 + (B_*^{\parallel}/H_{\parallel})^2]^{1/2}$.  
% Again, $B_* = \Phi_0/\pi\lambda_J d$ is the  parallel cross-over scale.  
but with a renormalized local stiffness $J_{\parallel}^{\prime}$
such that
$2\pi J_{\parallel}^{\prime}/k_B T = 1 + g_0^{\prime}/\pi v_F^{\prime\prime}$.
Comparison of the latter with Eq. (14) thus yields
$${2\pi J_{\parallel}^{\prime}\over{k_B T}} = 
1 + {v_F^{\prime\prime}\over{v_F^{\prime}}}
\Bigl({T_{dp}\over T} - 1\Bigr)\eqno (22)$$ 
for one over the effective  coupling constant
of the gaussian theory (21), where
$${v_F^{\prime\prime}\over{v_F^{\prime}}} =
\Biggl[1+\Biggl({B_*^{\parallel}\over{B_{\parallel}}}\Biggr)^2\Biggr]^{-1/2}
\eqno (23)$$ 
relates the Fermi velocity to the parallel field.

Above, we have reduced the pinned sine-Gordon model (12) to the
bosonic description of a Luttinger liquid with a single
backscattering impurity.$^{18}$
In particular, the equivalence (3) yields the  relationship
$Z_{\rm SG}\propto Z_{\rm LL}(U_{\perp})/Z_{\rm LL}(0)$
between the respective partion functions.  
The linear density
$\lambda_{flxn}^{-1} = \langle\partial_y\theta_-/2\pi i\rangle_{x_0}$
for half-loop (or fluxon$^7$) excitations
of Josephson vortices 
along the pin$^{28}$ is thus  given by the
difference 
$[\delta\, {\rm ln}\, Z_{\rm LL}/\delta (ia_0)]|_0^{U_{\perp}}$,
where the pinning term is held fixed.
Here, the field $a_0(y)$ is defined by
$\partial_y a_0 = V_{\rm KF}/v_F^{\prime\prime}$, where
$V_{\rm KF}  =
2\pi v_F^{\prime}\varepsilon_{\rm L}^{\parallel}/k_B T$
gives the voltage drop in the fermion analogy
(13). Kane and Fischer have computed
such  functional derivatives
using perturbation theory,$^{18}$
where they obtain   the result
$\delta\, {\rm ln}\, Z_{\rm LL}/\delta (ia_0)\propto
(V_{\rm KF}/V_0)^{\mu^{\prime}}$
with exponent
$\mu^{\prime} = 4\pi J_{\parallel}^{\prime}/k_B T - 1$.
Assuming that the voltage scale $V_0$ in the fermion analogy
is related to a (field dependent) current scale $I_0^{\prime}$
through the relationship
$V_{0}  =
2\pi v_F^{\prime}\varepsilon_{0}/k_B T$,
with $\varepsilon_{0} = c^{-1}(I_0^{\prime}/L_y)(\Phi_0/2\pi)$,
the above results lead to the relationship
$$\lambda_{flxn}^{-1} \propto 
  \Biggl({I\over{I_0^{\prime}}}\Biggr)^{\mu^{\prime}} -
\Biggl({I\over{I_0}}\Biggr)^{\mu}
\eqno (24)$$ 
for the density of fluxons along the pin,
with  
$$\mu = 2{T_{dp}\over T} - 1\eqno (25)$$
and  $I_0$ corresponding to the exponent
and to the current scale, respectively, in the limit of
infinite parallel field and/or Josephson penetration
length. 
%$\delta\mu = \mu^{\prime} - \mu$.  
%Comparison with Eq. (22) yields the expresion 
%$$\delta\mu\cong \Biggl({B_*^{\parallel}\over{B_{\parallel}}}\Biggr)^2
% \Biggl(1 - {T\over{T_{dp}}}\Biggr)\eqno (25)$$
%for the second exponent that is  valid in the limit of high
%fields for temperatures in the vicinity of the depinning
%transition. 
The assumption of flux
creep dynamics$^3$ then implies
that the true voltage drop between consecutive
layers is $V = (h/2e)(\bar c/\lambda_{flxn})$,
where $\bar c$ is the average creep velocity.
We therefore predict  that the
$c$-axis $I$-$V$ characteristic limited
by parallel collective pinning  is algebraic  at relatively
high parallel fields.   In particular, Eq. (24)
yields a voltage drop
$V\propto I^{\mu^{\prime}}$ for low currents $I\rightarrow 0$. 
%and that $V\propto \delta\mu I^{\mu} {\rm ln} (I_0/I)$ 
Also, both $\mu^{\prime}\rightarrow \mu$ and $I_0^{\prime}\rightarrow I_0$
as $B_{\parallel}\rightarrow \infty$.  Eq. (24) thus  indicates
that the system becomes more superconducting as parallel field increases. 
Finally, by Eq. (22) we have that $\mu = 1 = \mu^{\prime}$ at the depinning
transition, $T = T_{dp}$.  This means that ohmic
flux flow sets in at temperatures above or equal to  $T_{dp}$
by   Eq. (24).
%These results are summarized in Fig. 3.
Given the temperature dependence of
the parallel field scale $B_*^{\parallel} \sim H_1^{\parallel}$
(see Fig. 1),  the $c$-axis current should then  peak 
at some temperature $T_p < T_{dp}$ for fixed field and
voltage.

\bigskip
\bigskip
\centerline{\bf IV. Discussion}
\bigskip

The previous results evoke the following image for the parallel
thermodynamics of layered superconductors in the quasi-2D
regime:  Each Josephson vortex can be viewed as a string of
width $\Lambda_0$ confined to a given pair of consecutive
layers.  Parallel fluctuations of the string give rise to
an effective Josephson  penetration length, $\lambda_J (T) > \Lambda_0$,
as well as to entropic pressure.$^{14,15,24}$
The latter is responsible for both
the anomalous $H_{\parallel}^{-1}$ tail shown by the parallel
magnetization (11) and for the parallel collective pinning
effect discussed above.   
Although HTSC films that are equivalent to
$N$ Josephson-coupled layers (1)
already exist,    their physical properties have 
been examined only at temperatures far from the critical
region.$^{17,29}$  Comparable studies should be 
carried out within the critical regime of these materials to
test the predictions made here.

Finally, it must be stressed that all of the results obtained
here are only valid for physics at length scales {\it large} in comparison to
the ultraviolet cut-off $\alpha$ of the fermion analogy; e.g.,
for parallel fields $B_{\parallel} < \Phi_0/\alpha d$.  Clearly
the in-plane coherence length, $\xi_0$, which is roughly equal to the 
size of a typical Cooper pair, provides a lower bound for $\alpha$. 
In addition, the string image mentioned above suggests that
the bare Josephson penetration length $\Lambda_0$
supplies an upper bound for $\alpha$.  Where exactly 
within these limits $\alpha$ lies remains to be determined.

It is a pleasure to thank S. Sorella, G. Gomez-Santos, H. Safar,
M. Maley and U. Welp for very informative discussions.  This work
was supported in part by National
Science Foundation grant DMR-9322427 and by the Spanish Ministry for
Education and Culture. 

\vfill\eject
\centerline{\bf Appendix A: Derivation of Ginzburg-Landau Equations 
for Coupled}
\centerline{\bf Charge-Density Waves in One Dimension}
\bigskip

Below, we shall recover the  LD free-energy functional (1) from
the mean-field approximation (8) for the fermion analogy
[Eqs. (2a) - (6)] by extending Gorkov's original derivation of the 
Ginzburg-Landau field equation$^{25}$ to the case of  two coupled
CDW states in one dimension.  The limit
of low parallel field, $b_{\parallel} \ll \Lambda_0^{-1}$,
is assumed.
 
Consider two adjacent layers $l = 1, 2$ in isolation,
each  of length $L_y$ along the parallel field.
Assuming periodic boundary conditions in this direction,
we may then define the fictitious temperature
$T^{\prime} = v_F^{\prime}/L_y$ for the CDW state in
the fermion analogy.  Standard mean-field calculations
then yield that the  CDW is stable for fictitious temperatures
below a   critical temperature
$T_c^{\prime} = (e^{\gamma}/\pi)\Delta_0$, where
$\gamma$ denotes Euler's constant.  Note that the
thermodynamic limit $L_y\rightarrow\infty$ clearly
corresponds to  the  fictitious temperature
$T^{\prime} = 0$.  To begin the derivation, we first
observe that the mean-field equation
$H_l\Psi_l = \varepsilon_l\Psi_l$ plus the
definitions
$$\eqalignno {
G_l (x,t; x^{\prime},t^{\prime}) = &
-\langle T \Psi_R(x,l,t)\Psi_R^{\dag}(x^{\prime}, l, t^{\prime})\rangle 
&(A1)\cr
F_l(x,t; x^{\prime},t^{\prime}) = &
-\langle T \Psi_R^{\dag}(x,l,t)\Psi_L(x^{\prime}, l, t^{\prime})\rangle 
&(A2)\cr}$$
for the normal and the anomalous Greens functions, respectively,
ultimately lead to the Gorkov equations
$$\eqalignno {
(i\omega_n + i v_F\partial_x  + \mu_l) G_l (x,x^{\prime}, i\omega_n)
+\Delta_l(x) F_l^*(x,x^{\prime}, i\omega_n) & = \delta (x-x^{\prime}),
&(A3)\cr
(-i\omega_n - i v_F\partial_x + \mu_l)F_l^*(x,x^{\prime},i\omega_n) 
 - \Delta_l^*(x) G_l (x,x^{\prime},i\omega_n) & = 0, &(A4)\cr}$$
along with the associated gap equations
$$\eqalignno {
\Delta_1^*(x) & = - U_{\parallel} F_1^*(x,t; x,t)
- U_{\perp} F_2^*(x,t;x,t), & (A5)\cr
\Delta_2^*(x) & = -U_{\parallel} F_2^*(x,t;x,t) 
- U_{\perp} F_1^*(x,t;x,t). & (A6)\cr}$$
Here, $t = iy/v_F^{\prime}$ is the fictitious time variable,
while $i\omega_n$ denote the corresponding Matsubara frequencies.
Applying Gorkov's method$^{25}$ to these set of equations
yields the Ginzburg-Landau field equations
$$\eqalignno {
-(\partial_x - 2i k_{F,1})^2\Delta_1^* 
+ v_F^{-2}\Delta_{\sigma}^2(\Delta_1^*+\Delta_2^*) & = 0  &(A7)\cr
-(\partial_x - 2i k_{F,2})^2\Delta_2^*
+ v_F^{-2}\Delta_{\sigma}^2(\Delta_1^*+\Delta_2^*) & = 0  &(A8)\cr}$$
to lowest order in $\Delta_l$,
where
$$\Delta_{\sigma} = {4 e^{\gamma}\over{[7\zeta(3)]^{1/2}}}
\Biggl({2\pi v_F U_{\perp}\over{U_{\parallel}^2 - U_{\perp}^2}}\Biggr)^{1/2} 
\Delta_0 \eqno (A9)$$
is the pseudo-spin gap.  Above, $\zeta(z)$ denotes the
zeta function.    Finally, these equations can be
integrated, the result of which is the Ginzburg-Landau free-energy
functional
$$F\propto \int dx\Biggl\{ {1\over 2}
\Biggl({\partial\theta_1^{\prime}\over{\partial x}} - 2 k_{F,1}\Biggr)^2
+ {1\over 2}
\Biggl({\partial\theta_2^{\prime}\over{\partial x}} - 2 k_{F,2}\Biggr)^2 +
\lambda_J^{-2}[1 - {\rm cos}\,(\theta_1^{\prime} - \theta_2^{\prime})]\Biggr\}
\Delta_0^2, \eqno (A10)$$
for gaps of the form $\Delta_l (x) = (-1)^l\Delta_0 e^{i\theta_l^{\prime}(x)}$.
Above, $\lambda_J = v_F/|\Delta_{\sigma}|$ is the Josephson penetration
length.  After making the gauge transformation
$\theta_l^{\prime} (x) = \theta_l(x) + 2 k_{F,l} x$, we recover the
form (1) of the LD free-energy.  Last, it is worth mentioning that
$\lambda_J$ and $\Lambda_0$ are approximately equal along the
LE line, $T = 2\pi J_{\parallel}$, in  the
limit $|U_{\perp}|\ll |U_{\parallel}|$.  This is demonstrated by
observing that we have
$U_{\parallel} = - 6\pi v_F/5$ 
and 
$\Lambda_0 = (\pi v_F^{\prime}/|U_{\perp}|)^{1/2} \alpha$
in such case [see Eqs. (6)], and by substitution of the former  
into Eq. (A9).  On the other hand, 
Eq. (A9) also indicates that $\lambda_J$ diverges exponentially
as $T$ approaches the decoupling temperature $T_*$ from below, since 
$\Delta_0\cong 2\omega_0{\rm exp}[-2\pi v_F/(2U_{\perp} - U_{\parallel})]$.
This agrees with results based on the Coulomb gas analogy$^{14}$ for
the LD model (1), as well as with those based on a model for ``frozen''
layered superconductors in the Meissner phase.$^{24}$

\vfill\eject
\centerline{\bf Appendix B: Derivation of 
Effective Massless Thirring Model}
%\centerline{\bf Charge-Density Waves in One Dimension}
\bigskip

Below, we shall obtain the effective low-energy
Hamiltonian (20) of  the massive Thirring model
analogy (13) for a pinned double-layer superconductor (12) in
the presence of parallel field.
The limit of a thick pinning plane
$$\xi_p > a_0,\eqno (B1)$$
with respect to the average  
separation in between Josephson vortices (fermions)
is assumed to insure the validity of the Luttinger-Liquid hypothesis 
% with respect to the pinning-related terms 
in the fermion analogy (13).
Note that the above separation is related to 
the parallel magnetic induction and to the Fermi wavenumber by
$b_{\parallel} = 2\pi/a_0 = 2 k_F$.  The 
%passage into the  
gapless Luttinger model will be achieved 
%effected 
through the canonical transformation
[see Eqs. (15) - (19)] of
the original left and right moving fields
$$\eqalignno{
\Psi_R (x) & =  L_x^{-1/2}\sum_k (x_k c_k - y_k d_k) e^{ikx} &(B2)\cr
\Psi_L (x) & = L_x^{-1/2}\sum_k (y_k c_k + x_k d_k) e^{ikx} &(B3)\cr}$$
into the new fields
$\Psi_+ (x)  =  L_x^{-1/2}\sum_k c_k e^{ikx}$ and 
$\Psi_- (x)  =  L_x^{-1/2}\sum_k d_k e^{ikx}$. 
We now begin the derivation of Eq. (20) by 
applying this canonical transformation to each term in the
fermion analogy (13).

{\it Kinetic Energy.}  The kinetic energy is given by
$H_0 = \sum_k (\varepsilon_k^{+} c_k^{\dag} c_k
+ \varepsilon_k^{-} d_k^{\dag} d_k)$, where
$\varepsilon_k^{\pm} = \pm \theta (k) E_k \mp \theta (-k) E_k$
are the energy eigenvalues of the quasi-particle excitations.
Since only those  excitations that are   near the Fermi
surface are relevant, we can approximate the quasi-particle
energy spectrum by
$\varepsilon_k^{\pm} \cong  \varepsilon_F \pm  v_F^{\prime\prime}
(k\mp k_F)$,
where $v_F^{\prime\prime} = (v_F^{\prime} k_F/E_{k_F}) v_F^{\prime}$
is the group velocity at the Fermi surface.  
This immediately yields the new expression 
$$H_0 = \int dx v_F^{\prime\prime}
(\Psi_-^{\dag} i\partial_x \Psi_- - \Psi_+^{\dag} i\partial_x \Psi_+)
\eqno (B4)$$
for the kinetic energy modulo a trivial
shift of the chemical potential.

{\it Potential Energy.}  Using     the form
$V_{\rm KF} (x) = \sum_q e^{iqx} V_{\rm KF}(q)$ for potential
energy drop at $x = x_0$ in terms of its Fourier
transform $V_{\rm KF}(q)$, we can reexpress the corresponding
term, $H_{\rm KF} = 2^{1/2} \int dx  V_{\rm KF}
\Psi^{\dag}\Psi$,
 in the fermion analogy by
$$\eqalignno{
H_{\rm KF} = 2^{1/2} \sum_{k,q} & V_{\rm KF}(q) 
[(x_{k+q} x_k + y_{k+q} y_k) c_{k+q}^{\dag} c_k
- (x_{k+q} y_k - y_{k+q} x_k) c_{k+q}^{\dag} d_k\cr
  & - (y_{k+q} x_k - x_{k+q} y_k) d_{k+q}^{\dag} c_k
    + (y_{k+q} y_k + x_{k+q} x_k) d_{k+q}^{\dag} d_k].& (B5)\cr}$$
After making the Luttinger-Liquid-type approximation
$(x_{k+q}, y_{k+q})\rightarrow (x_k, y_k)$ above
for the coherence factors, which is valid in the limit
(B1) $k_F\xi_p\gg 1$, we recover the original simple form
$$H_{\rm KF} \cong 2^{1/2} \int dx V_{\rm KF}
(\Psi_+^{\dag}\Psi_+ + \Psi_-^{\dag}\Psi_-) \eqno (B6)$$
for the potential energy
in terms of the new fields. Here,
we have used the identity $x_k^2 + y_k^2 =1$.

{\it Pinning/Backscattering.}  Let us first reexpress the pinning energy
(13) as
$$H_{\rm pin} = - \Delta_{\sigma}\int_{x_0 -\xi_p/2}^{x_0 + \xi_p/2} dx
(\Psi_L^{\dag}\Psi_R + \Psi_R^{\dag}\Psi_L).
\eqno (B7)$$
If we now  take  the long-wavelength limit (B1)
$k_F\xi_p\gg 1$, then the bounds on the above  integral can be
extended to $\pm\infty$.  After Fourier transformation, we obtain
the expression
$$H_{\rm pin} \rightarrow
- \Delta_{\sigma} \sum_k[(x_k^2 - y_k^2)(d_k^{\dag} c_k + c_k^{\dag} d_k)
+ 2 x_k y_k(c_k^{\dag} c_k - d_k^{\dag} d_k)]\eqno (B8)$$
as a result.
If we then make the additional replacements
$$x_k\rightarrow u_{k_F}\quad {\rm and}\quad
y_k\rightarrow ({\rm sgn}\, k) v_{k_F}\eqno (B9)$$ 
%$$(x_k, y_k)\rightarrow 
%\theta(k)(x_{k_F}, y_{k_F}) + \theta (-k) (x_{-k_F}, y_{-k_F})$$
valid under the Luttinger-Liquid hypothesis, tracing back
the previous steps leads to the effective low-energy Hamiltonian 
$$H_{\rm pin}\cong - (u_{k_F}^2 - v_{k_F}^2)\xi_p\Delta_{\sigma}
(\Psi_-^{\dag}\Psi_+ + \Psi_+^{\dag}\Psi_-)|_{x_0}
- 2 u_{k_F} v_{k_F} \xi_p\Delta_{\sigma}
(\Psi_+^{\dag}\Psi_+ + \Psi_-^{\dag}\Psi_-)|_{x_0}
\eqno (B10)$$
corresponding to the pinning term.
Last, 
%because the number of left and right moving fermions
%are equal, 
the second term above represents a trivial shift
of the chemical potential in the long
wave-length limit (B1).  The effective pinning term 
at the Fermi surface  thus has the same form as the original, but
with a renormalized pinning scale
$\xi_p^{\prime} = (v_F^{\prime} k_F/E_{k_F})\xi_p$.

{\it Interaction.}  Substituting the canonical transformation
(B2) and (B3) into the forward scattering interaction
term $H_1 = 2g_0\int dx \Psi_L^{\dag} \Psi_R^{\dag}
\Psi_R \Psi_L$ yields an  equivalent expression 
of the form
$$\eqalignno{
H_1 = {2g_0\over{L_x}}\sum_{k,k^{\prime},q} &
(y_{k+q} y_k c_{k+q}^{\dag} c_k + y_{k+q} x_k c_{k+q}^{\dag} d_k
+x_{k+q} y_k d_{k+q}^{\dag} c_k + x_{k+q} x_k d_{k+q}^{\dag} d_k)\times\cr
& \times 
(x_{k^{\prime}-q} x_{k^{\prime}} c_{k^{\prime}-q}^{\dag} c_{k^{\prime}} 
- x_{k^{\prime}-q} y_{k^{\prime}} c_{k^{\prime}-q}^{\dag} d_{k^{\prime}}
- y_{k^{\prime}-q} x_{k^{\prime}} d_{k^{\prime}-q}^{\dag} c_{k^{\prime}} 
+ y_{k^{\prime}-q} y_{k^{\prime}} d_{k^{\prime}-q}^{\dag} d_{k^{\prime}}).&\cr
&&(B11)\cr}$$
After making the replacements (B9)
in the coherence factors above, 
which is  valid for excitations near the Fermi surface, 
%such that $|q|\ll k_F$, 
we recover the original form
$$H_1 \cong  
%{2g_0\over{L_x}} \sum_{k,k^{\prime},q} d_{k+q}^{\dag} d_k 
%c_{k^{\prime}-q}^{\dag} c_{k^{\prime}} =
2g_0 \Biggl({v_F^{\prime} k_F\over{E_{k_F}}}\Biggr)^2
\int dx \Psi_-^{\dag} \Psi_+^{\dag} 
\Psi_+ \Psi_- + H_4\eqno (B12)$$
for the interaction energy in terms of the new fields,
%In so doing we employ the identity $x_k^4 + 2 x_k^2 y_k^2+ y_k^4 = 1$.
in addition to a forward scattering contribution
$$H_4 =  2g_0 (u_{k_F} v_{k_F})^2
\sum_q [\rho_+(q)\rho_+(-q) + \rho_-(q) \rho_-(-q)]\eqno (B13)$$
that appears in terms of the new
particle-hole operators 
$\rho_+(q) = L_x^{-1/2} \sum_k c_{k+q}^{\dag} c_k$
and $\rho_-(q) = L_x^{-1/2} \sum_k d_{k+q}^{\dag} d_k$
for right and left moving fermions.
This interaction can be incorporated into the kinetic energy
(B4) via Kronig's identity,$^{20}$ which yields the final result
$$v_F^{\prime\prime} = v_F^{\prime}
\Biggl({v_F^{\prime} k_F\over{E_{k_F}}} + {g_0\over{2\pi v_F^{\prime}}}
{\Delta_{\sigma}^2\over{E_{k_F}^2}}\Biggr)\eqno (B14)$$
for the effective Fermi velocity.  Notice, however, that this correction
is negligible in the limit of high parallel fields, 
$\Delta_{\sigma}\ll E_{k_F}$, and at temperatures near the depinning
transition, $|g_0| \ll \pi v_F^{\prime}$.

\vfill\eject
\centerline{\bf References}
\vskip 16 pt

\item {1.}  See for example {\it The Physical Properties of
High-Temperature Superconductors}, edited by
D.M. Ginsberg (World Scientific, Singapore, 1989).

\item {2.}  V. Ambegaokar and A. Baratoff, Phys. Rev. Lett.
{\bf 10}, 486 (1963); (E) {\bf 11}, 104 (1963);
W.E. Lawrence and S. Doniach, in
{\it Proceedings of the  12$^{\rm th}$
International Conference on  Low Temperature
Physics (Kyoto, 1970)}, edited by E. Kanda
(Keigaku, Tokyo, 1971) p. 361. 

\item {3.}  G. Blatter, M.V. Feigel'man, V.B. Geshkenbein, A.I. Larkin,
and V.M. Vinokur, Rev. Mod. Phys. {\rm 66}, 1125 (1994). 

\item {4.} J. Friedel, J. Phys. (Paris) {\bf 49}, 1561 (1988).

\item {5.} S.N. Artemenko, I.G. Gorlova, and Yu. I. Latyshev, 
Phys. Lett. A {\bf 138}, 428 (1989).

\item {6.} A. Junod, in {\it Studies of High Temperature 
Superconductors}, edited by A.V. Narlikar
(Nova Science, New York, 1996), Vol. 18.

\item {7.}  S.E. Korshunov, Europhys. Lett. {\bf 11}, 757 (1990).

\item {8.} B. Horovitz, Phys. Rev. Lett. {\bf 72}, 1569 (1994);
Phys. Rev. B {\bf 47}, 5964 (1993).

\item {9.} J.P. Rodriguez, Europhys. Lett. {\bf 31}, 479 (1995).

\item {10.} Y.M. Wan, S.E. Hebboul, D.C. Harris, and J.C. Garland,
Phys. Rev. Lett. {\bf 71}, 157 (1993); {\bf 74}, 5286 (E) (1995);
Y.M. Wan, S.E. Hebboul, and J.C. Garland, Phys. Rev. Lett.
{\bf 72}, 3867 (1994).

\item {11.} M. Friesen, Phys. Rev. B {\bf 51}, 12786 (1995);
S.W. Pierson, Phys. Rev. Lett. {\bf 75}, 4674 (1995);
see also D.I. Glazman and A.E. Koshelev, Zh. Eksp.
Teor. Fiz. {\bf 97}, 1371 (1990)
[Sov. Phys. JETP {\bf 70}, 774 (1990)].

\item {12.} L.I. Glazman and A.E. Koshelev, Phys. Rev. B {\bf 43}, 2835 (1991).

\item {13.} Y. Matsuda, M.B. Gaifullin, K. Kumagai, K. Kadowaki,
and T. Mochiku, Phys. Rev. Lett. {\bf 75}, 4512 (1995).

\item {14.} J.P. Rodriguez, 
%``On the Decoupling of Layered Superconducting
%Films in Parallel Magnetic Field'', 
%Los Alamos preprint LA-UR-95-1908
J. Phys. Cond. Matter {\bf 9}, 5117 (1997). (cond-mat/9604182).

\item {15.} J.P. Rodriguez, 
%``Fermion Analogy for Layered
%Superconducting Films in Parallel Magnetic Field'', 
%ICMM-CSIC report (1995)
Europhys. Lett. {\bf 39}, 195 (1997). (cond-mat/9606154).
[Other analogies employ fermions that
live {\it in between} consecutive layers (see ref. 8).]

\item {16.} L. Bulaevskii and J.R. Clem, Phys. Rev. B {\bf 44}, 10234 (1991).

\item {17.} M. Oussena, P.A.J. de Groot, R. Gagnon, and L. 
Taillefer, Phys. Rev. Lett. {\bf 72}, 3606 (1994).

\item {18.} C.L. Kane and M.P.A. Fisher, Phys. Rev. B {\bf 46},
15233 (1992).

\item {19.} A. Luther and V.J. Emery, Phys. Rev. Lett. {\bf 33}, 589
(1974); P.A. Lee, Phys. Rev. Lett. {\bf 34}, 1247 (1975).

\item {20.} J. Voit, Rep. Prog. Phys. {\bf 58}, 977 (1995).

\item {21.} S. Coleman, Phys. Rev. D {\bf 11}, 2088 (1975).

\item {22.}  The mean-field theory, (7) and (8), respects
the following local gauge invariance:
$\Psi_l\rightarrow e^{i\alpha\sigma_3}\Psi_l$,
$\Delta_l\rightarrow e^{2\alpha i}\Delta_l$, and
$\mu_l\rightarrow\mu_l - v_F \partial_x\alpha$,
where $\alpha (x)$ is an arbitrary function.

\item {23.} S.T. Chui and P.A. Lee, Phys. Rev. Lett. {\bf 35}, 315 (1975).

\item {24.} J.P. Rodriguez, Phys. Rev. B {\bf 54}, 497 (1996).

\item {25.} L.P. Gorkov, Sov. Phys.-JETP {\bf 9}, 1364 (1959);
A.A. Abrikosov, L.P. Gorkov, and I.E.
Dzyaloshinski, {\it Methods of Quantum Field Theory in
Statistical Physics} (Dover, New York, 1975).

%\item {26.} J.P. Rodriguez, unpublished.

\item {26.} E. Zeldov, D. Majer, M. Konczykowski,
V.B. Geshkenbein, V.M. Vinokur, and H. Shtrikman,
Nature {\bf 375}, 373 (1995).

\item {27.}  R.P. Huebener and E.H. Brandt,
Physica C {\bf 202}, 321 (1992).

\item {28.} D.R. Nelson and V.M. Vinokur, Phys. Rev. B {\bf 48}, 13060 (1993).

\item {29.} G. Hechtfischer, R. Kleiner, A.V. Ustinov, and P. M\"uller,
Phys. Rev. Lett. {\bf 79}, 1365 (1997), and references therein.

\vfill\eject
\centerline{\bf Figure Caption}
\vskip 20pt
\item {Fig. 1.}  Shown is the phase diagram for 
thin-film layered superconductors
in the quasi-2D regime as a function of the parallel field, $H_{\parallel}$.
The integer $l_1$ designates a parallel vortex lattice in which
flux penetrates every $l_1$ layers (see ref. 16).  
Also, $T_{c0}$ denotes the
mean-field transition temperature of an isolated layer.

\item {Fig. 2.}  The equilibrium parallel magnetization
[Eq. (11)]
obtained from the mean-field theory approximation
to the  fermion analogy  is displayed in the
vicinity of the decoupling transition, $T\lsim T_*$ and/or
$B_{\perp}\gsim \Phi_0/\lambda_J^2$.

\end

\item {4.} K.B. Efetov, Zh. Eksp. Teor. Fiz. {\bf 76}, 1781 (1979)
[Sov. Phys. JETP {\bf 49}, 905 (1979)].

\item {6.} L.V. Mikheev and E.B. Kolomeisky, Phys. Rev. B {\bf 43},
10431 (1991) (Both this ref. and the following one discusses
an analogy for the LD model
in terms of fermions that
exist  in {\it between} consecutive layers,
which is contrary to the present
work where the fermions live  {\it on} each layer).

\item {8.} S.E. Korshunov and A.I. Larkin, Phys. Rev. B {\bf 46},
6395 (1992).

\item {11.} Y.H. Li and S. Teitel, Phys. Rev. B {\bf 47}, 359
(1993); {\it ibid} {\bf 49}, 4136 (1994).

\item {12.} S. Hikami and T. Tsuneto, Prog. Theor. Phys. {\bf 63},
387 (1980); B. Chattopadyay and S. R. Shenoy,
Phys. Rev. Lett. {\bf 72}, 400 (1994).

\item {15.} A.M. Polyakov, Nucl. Phys. B{\bf 120}, 429 (1977);
{\it Gauge Fields and Strings} (Harwood, New York, 1987).

\item {21.}  A result similar to Eq. (10) was obtained long ago by
V.L. Pokrovskii and A.L. Talanov, Zh. Eksp. Teor. Fiz. {\bf 78},
269 (1980) [Sov. Phys. JETP {\bf 51}, 134 (1980)], for the case of
2D incommensurate crystals.
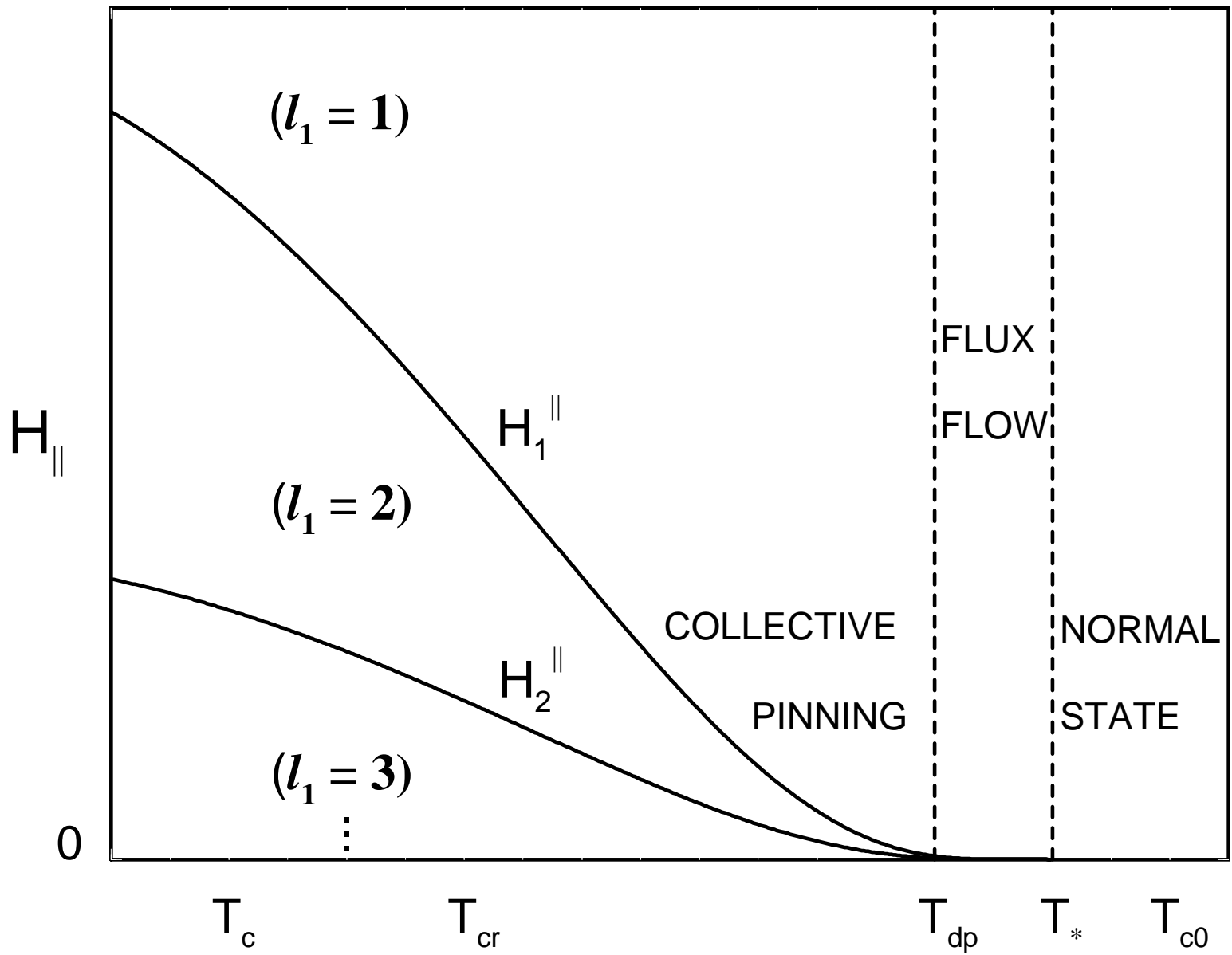
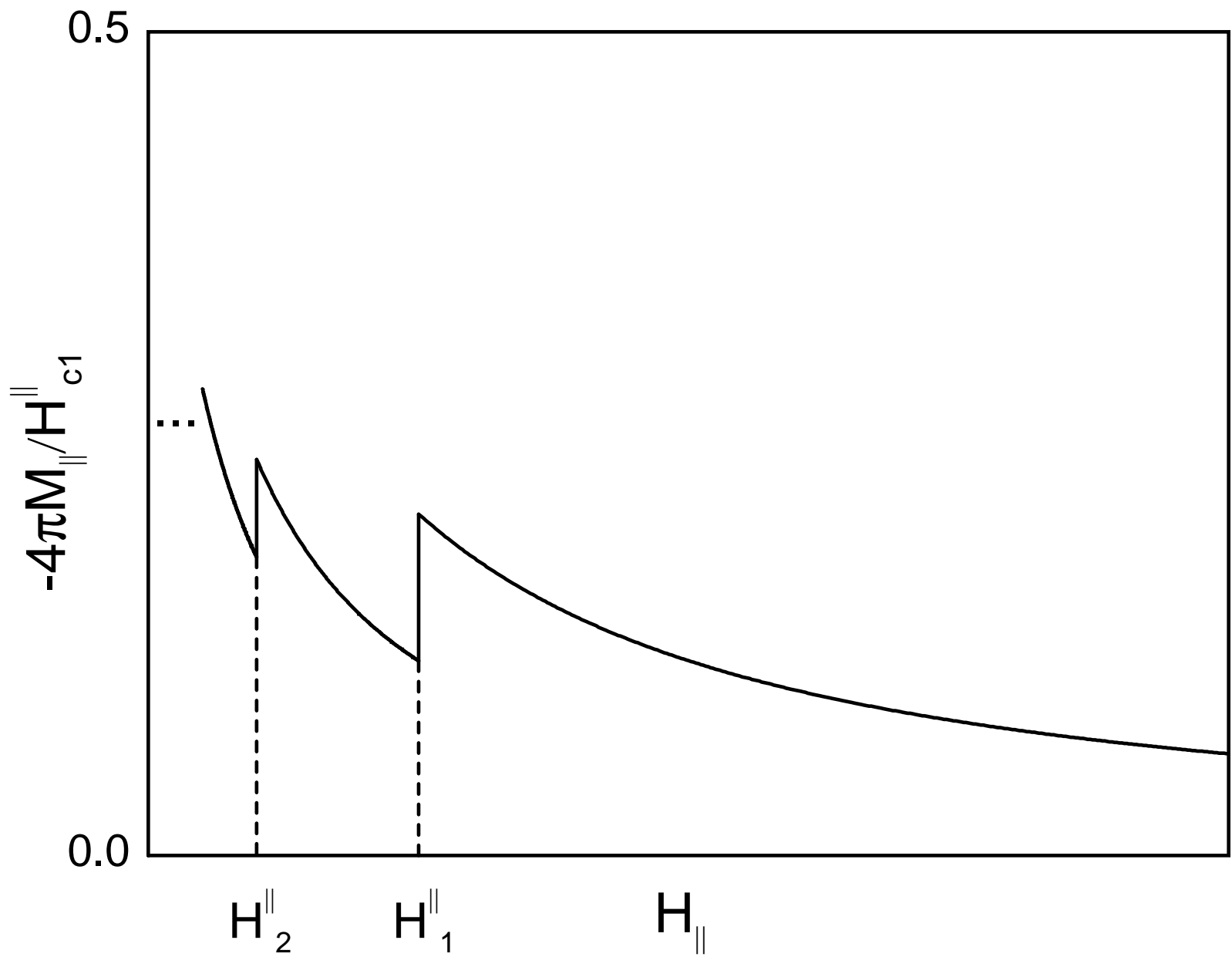
